\documentclass[apjl]{emulateapj}
\slugcomment{(To appear in the ApJ Letters)}
\usepackage[bookmarks=false, colorlinks=true, citecolor=blue, backref=true]{hyperref}  
\newcommand{\um}{\mbox{$\,\mu{\rm m}$}}

\newcommand{\etal}{et al.~}
\newcommand{\kms} {\mbox{\,km~s$^{-1}$}}

\newcommand{\LFIR}{\mbox{$L_{\rm FIR}$}}

\newcommand{\NII}{\mbox{[N\,{\sc ii}]}}
\newcommand{\CII}{\mbox{[C\,{\sc ii}]}}

\begin{document}

\shorttitle{\NII\ 205\um\ Emission in BRI\,1202-0725}
\shortauthors{Lu et al.}

\title{ALMA \NII\,205\um\ Imaging Spectroscopy of the Interacting Galaxy System \\
		BRI\,1202-0725 at Redshift 4.7\footnotemark[$\star$]}

\author{
Nanyao Lu\altaffilmark{1,2},
Yinghe Zhao\altaffilmark{3,4,5},
Tanio D\'iaz-Santos\altaffilmark{6},
C. Kevin Xu\altaffilmark{1,2},  
Vassilis Charmandaris\altaffilmark{7,8}, 
Yu Gao\altaffilmark{9},
Paul P. van der Werf\altaffilmark{10},
George C. Privon\altaffilmark{11,12},
Hanae Inami\altaffilmark{13},
Dimitra Rigopoulou\altaffilmark{14},
David B. Sanders\altaffilmark{15},
Lei Zhu\altaffilmark{1,2}
}
\altaffiltext{1}{National Astronomical Observatories, Chinese Academy of Sciences (CAS), Beijing 100012, China; nanyao.lu@gmail.com}
\altaffiltext{2}{China-Chile Joint Center for Astronomy, Camino El Observatorio 1515, Las Condes, Santiago, Chile}
\altaffiltext{3}{Yunnan Observatories, Chinese Academy of Sciences, Kun- ming 650011, China}
\altaffiltext{4}{Key Laboratory for the Structure and Evolution of Celestial Objects, Chinese Academy of Sciences, Kunming 650011, China}
\altaffiltext{5}{Center for Astronomical Mega-Science, CAS, 20A Datun Road, Chaoyang District, Beijing 100012, China}
\altaffiltext{6}{Nucleo de Astronomia de la Facultad de Ingenieria, Universidad Diego Portales, 
                 Av. Ejercito Libertador 441, Santiago, Chile}
\altaffiltext{7}{Department of Physics, University of Crete, GR-71003 Heraklion, Greece}
\altaffiltext{8}{IAASARS, National Observatory of Athens, GR-15236, Penteli, Greece}
\altaffiltext{9}{Purple Mountain Observatory, CAS, Nanjing 210008, China}
\altaffiltext{10}{Leiden Observatory, Leiden University, PO Box 9513, 2300 RA Leiden, The Netherlands}
\altaffiltext{11}{Departamento de Astronom\'ia, Universidad de Concepci\'on, Casilla 160-C, Concepci\'on, Chile}
\altaffiltext{12}{Pontificia Universidad Cat\'olica de Chile, Instituto de Astrofisica, Casilla 306, Santiago 22, Chile}
\altaffiltext{13}{Centre de Recherche Astrophysique de Lyon (CRAL), Observatoire de Lyon, CNRS, UMR5574, F-69230, Saint-Genis-Laval, France}
\altaffiltext{14}{Department of Physics, University of Oxford, Keble Road, Oxford OX1 3RH, UK}
\altaffiltext{15}{University of Hawaii, Institute for Astronomy, 2680 Woodlawn Drive, Honolulu, HI 96822, USA}

\footnotetext[$\star$]{
The National Radio Astronomy Observatory is a facility of the National Science Foundation operated under cooperative  
agreement by Associated Universities, Inc.}


\begin{abstract}
\noindent
We present the results from Atacama Large Millimeter/submillimeter Array 
(ALMA) imaging in the \NII\ 205\um\ fine-structure line (hereafter \NII) 
and the underlying continuum of BRI\,1202-0725, an interacting galaxy 
system at $z =$ 4.7, consisting of an optical QSO, a sub-millimeter galaxy
(SMG) and two Lyman-$\alpha$ emitters (LAEs), all within $\sim$25\,kpc 
of the QSO. We detect the QSO and SMG in both \NII\ and continuum.  At 
the $\sim$1\arcsec\ (or 6.6 kpc) resolution, both QSO and SMG are resolved
in \NII, with the de-convolved major axes of $\sim$9 and $\sim$14\,kpc, 
respectively.  In contrast, their continuum emissions are much more 
compact and unresolved even at an enhanced resolution of $\sim$0.7\arcsec.
The ratio of the \NII\ flux to the existing CO\,(7$-$6) flux is used to
constrain the dust temperature ($T_{\rm dust}$) for a more accurate 
determination of the FIR luminosity \LFIR. Our best estimated $T_{\rm dust}$
equals $43\ (\pm 2)$ K for both galaxies (assuming an emissivity index 
$\beta = 1.8$).  The resulting $L_{\rm CO(7-6)}/$\LFIR\ ratios are 
statistically consistent with that of local luminous infrared galaxies,
confirming that $L_{\rm CO(7-6)}$ traces the star formation (SF) rate (SFR)
in these galaxies.  We estimate that the on-going SF of the QSO (SMG) has 
a SFR of 5.1 $(6.9) \times 10^3\,M_{\odot}$\,yr$^{-1}$ ($\pm$ 30\%) 
assuming Chabrier initial mass function, takes place within a diameter
(at half maximum) of 1.3 (1.5)\,kpc, and shall consume the existing 5 $(5) 
\times 10^{11}\,M_{\odot}$ of molecular gas in 10 $(7) \times 10^7$ years. 
\end{abstract}

\setcounter{footnote}{2}

\section{INTRODUCTION} \label{sec1}

Star formation (SF) rate (SFR) measures the fundamental physical process
of transforming gas into stars and is one of the most important drivers 
of galaxy evolution.  For high-$z$ galaxies,  Lu et al. (2015; hereafter 
Lu15) explored a new dual-spectral line approach for estimating {\it both}
SFR and the far-infrared (FIR) color $C(60/100)$ (thus $T_{\rm dust}$), 
where $C(60/100)$ refers to the rest-frame $f_{\nu}(60\um)/f_{\nu}(100\um)$
ratio.  For local (ultra-)luminous infrared galaxies 
[(U)LIRGs], the luminosity of the CO\,(7$-$6) line emission,
$L_{\rm CO(7-6)}$, can be used to infer the SFR of the galaxy with 
a $\sim$30\% accuracy, irrespective of whether the galaxy hosts an 
active galactic nucleus (AGN) (Lu et al. 2014, 2017; Lu15). Furthermore,
by measuring the flux of the \NII\ 205\um\ line (1461.13 GHz; 
hereafter as [NII]), one can use the steep dependence of the \NII\ 
to CO\,(7$-$6) flux ratio on $C(60/100)$ to estimate $C(60/100)$
or $T_{\rm dust}$, with an accuracy equivalent to $\sim$2\,K in 
$T_{\rm dust}$ if the dust emissivity power law index $\beta$ is 
around 2 (Lu15).  $C(60/100)$ is empirically related to $\Sigma_{\rm SFR}$,
the average SFR surface density (Liu et al. 2015; Lutz et al. 2016),
another fundamental parameter of galaxy SF.  This indirect estimate
of $\Sigma_{\rm SFR}$ is useful at high $z$, where it is often challenging
to spatially resolve a galaxy.  This dual-line strategy also allows 
for estimating additional galaxy physical parameters, including 
the SF area ($\approx$ SFR/$\Sigma_{\rm SFR}$), the molecular gas 
mass ($M_{\rm gas}$) from the continuum flux underlying the CO\,(7$-$6)
line (Scoville et al. 2016), and the gas depletion time $\tau_{\rm gas}$
($= M_{\rm gas}$/SFR). If the lines are sufficiently resolved spectrally
and spatially, insights into the gas dynamics can also be gained.

In a Cycle-3 program with the Atacama Large Millimeter/submillimeter 
Array (ALMA), we conducted a spectral line snapshot survey of 8 ULIRGs
and 4 LIRGs of $4 < z \lesssim 5.5$ to complete their detections in 
\NII\ and CO\,(7$-$6).  All our targets have prior detections in 
the \CII\ 158\um\ line (hereafter \CII).  In this paper we present
the results from our \NII\ observation of the interacting galaxy group
BRI\,1202-0725 at $z = 4.7$ (Isaak et al. 1994).  This un-lensed
system consists of 2 ULIRGs: a QSO at $z = 4.695$ and an optically 
obscured, sub-millimeter galaxy (SMG) at $z = 4.692$ and 3.8\arcsec\ 
($\sim$25 kpc) northwest of the QSO (Omont et al. 1996; 
Ohta et al. 1996; Hu et al. 1996; Yun et al. 2000; Carilli et al. 2002;
Yun \& Carilli 2002; Momjian et al. 2005; Iono et al. 2006; Salom\'e
et al. 2012).  In addition, two Lyman-$\alpha$ emitting galaxies 
(LAEs) have been detected in this system, with the first one (LAE1)
located between the QSO and SMG, at $\sim$2.3\arcsec\ northwest of
the QSO and the second one (LAE2) at 2.7\arcsec\ southwest of the QSO
(Hu et al. 1996; Fontana et al. 1996; Ohta et al. 2000; Ohyama et al. 2004).

The \CII\ emission was detected with ALMA on all the member galaxies,
along with a possible extended gas ridge between the QSO and SMG 
(Wagg et al. 2012; Carilli et al. 2013).  Both QSO and SMG were 
also detected in CO\,(7$-$6) and CO\,(5$-$4) (Omont et al. 1996;
Salom\'e et al. 2012).  Decarli et al. (2014) presented 
the \NII\ spectra of the system obtained with the IRAM interferometer
and suggested possible \NII\ detections for the SMG and LAEs,
but a non-detection for the QSO.  However, the \NII\ line peak flux
densities of both QSO and SMG (see Fig.~2 here) are either near 
or below the spectral noise in Decarli et al.  This might explain 
the flux discrepancy between ours and theirs.  Pavesi et al. (2016) 
analyzed an archival ALMA observation and derived the \NII\ fluxes 
of 0.74 ($\pm$0.07), 1.5 ($\pm$0.2) and 0.30 ($\pm$0.06) Jy\,\kms\ 
for the QSO, SMG and LAE2, respectively, and a 3$\sigma$ flux density 
upper limit of $0.5$\,mJy for LAE1.
 
The \NII\ line is a major cooling line for ionized gas in galaxies.
This line has only been detected in a handful of galaxies at $z 
\gtrsim 4$ (e.g., Combes et al. 2012; Decarli et al. 2012; Nagao 
et al. 2012; Rawle et al. 2014; B\'ethermin et al 2016;  Pavesi 
et al. 2016).  Our ALMA \NII\ observation of BRI 1202-0725 detected 
both SMG and QSO at good S/N ratios.  In \S2 we describe our 
observations and results.  In \S3 we analyze the observed \NII\ and
dust continuum emissions, and discuss the SF properties and \NII/CII\
flux ratios of the galaxies in BRI 1202-0725.  We use a flat 
cosmology with $\Omega_M = 0.27$, $\Omega_{\Lambda} = 0.73$ and $H_0
= 71$\,\kms\,Mpc$^{-1}$. At $z = 4.694$, the luminosity distance is 
44,172 Mpc and 1\arcsec\ corresponds to 6.6 kpc.
\vspace{0.1in}

\section{Observations and Results} \label{sec2}

BRI 1202-0725 was observed in ALMA Band 6 in the time division 
mode.  Of the 4 spectral windows (SPWs), each of 1875 MHz wide, 
one was targeted at the redshifted \NII\ line (at $\sim$256.6 GHz) 
and the other 3, centered at 254.6, 240.6 and 238.6 GHz,
respectively, were used for continuum measurements.  Each SPW has 
128 channels with an effective resolution of 31.25 MHz ($\sim$36.5\kms).  
The observation consisted of 2 independent executions, each 
with a 20.3 min on-target integration.  The first was executed 
on January 8, 2016 using 37 antennas covering baselines from 
15.1 to 310.2 meters and the second on March 13, 2016 using 38
antennas and baselines ranging from 15.1 to 460.0 meters.  
The phase, bandpass and flux calibrations were based on the quarsars
J1229+0230 and J1159-0940.  The data reduction was carried out 
with the Common Astronomy Software Applications (CASA) 4.5.3 
and the final images were cleaned using the natural weighting,
resulting in a synthesized beam of 1.0\arcsec$\times$0.8\arcsec\ 
(full width at half maximum; FWHM) at a position angle (PA) of 
75\arcdeg\ (N to E) for both continuum and line data.  
The r.m.s.~noise in the continuum image, which is the average 
of the 3 continuum SPWs, is $\sim$34\,$\mu$Jy\,beam$^{-1}$. 
The final \NII\ spectral cube has a velocity channel width of 
100\,\kms, with an r.m.s. noise of $\sim$0.3\,mJy\,beam$^{-1}$
in individual channels.

We compare in Fig.~1 the continuum image and the \NII\ image
integrated over $\nu_{obs} =$ 256.148 to 257.090 GHz, which 
encompasses all and the vast majority of the \NII\ fluxes 
from QSO and SMG, respectively.   While the QSO, SMG and LAE2 
are all detected in the continuum, only the QSO and SMG are 
clearly detected in \NII.  There appears to exist some faint 
(up to 3$\sigma$ or 0.1 mJy\,beam$^{-1}$) dust emission 
connecting the QSO and SMG, with a morphology similar to 
what is seen in \CII\ (Carilli et al. 2013). However, its 
counterpart is not detected in the \NII\ image.  The outer \NII\
emission contours (up to 3$\sigma$ levels) of both QSO and SMG 
show distortions. To gain a better insight into this, we 
also cleaned the data using the Briggs weighting (with {\it 
robust} $= 0$) to lower the sidelobes and to effectively enhance
the resolution to 0.7\arcsec$\times$0.6\arcsec\ (PA 
$\approx$ 78\arcdeg).  The results are shown in the insert in 
each panel in Fig. 1 for an image section covering our targets. 
While the \NII\ emission of the SMG remains in a similar morphology, 
that of the QSO is marginally resolved into 3 peaks: (i) the brightest
one aligns with the peak dust emission; (ii) the second one
(12$^{\rm h}$05$^{\rm m}$23{\fs}09, -7{\arcdeg}42{\arcmin}33{\farcs}2)
is $\sim$0.7\arcsec\ southwest of (i), in the direction of LAE2;
(iii) the third one (12$^{\rm h}$05$^{\rm m}$23{\fs}06, 
-7{\arcdeg}42{\arcmin}32{\farcs}5) is $\sim$1\arcsec\ northwest 
of (i), with a peak surface brightness just under 3$\sigma$.  
While (i) and (iii) have similar line central velocities, (ii) 
is blue-shifted by $\sim$200\,\kms\ relative to (i).
Contributing to $\sim$10\% of the total \NII\ flux of the QSO, 
(iii) is the main cause of the distorted contours seen. The relative
positions of (ii) and (iii) are suggestive of tidal interaction or gas 
connection between the member galaxies.

\section{Analysis and Discussion} \label{sec3}

In this section we analyze the \NII\ and continuum emissions 
and derive the parameters in Table~1 using the data cleaned 
with the natural weighting.

\subsection{{\rm \NII} Line Emission} \label{sec3.1}

In Fig.~2 we show the \NII\ spectra extracted within 
the elliptical apertures defined in Table 1.  For the QSO 
and SMG, we fit to their spectrum a single Gaussian profile
plus a constant to derive the line flux. The resulting \NII\ 
line widths are similar to or slightly wider than those of
the corresponding CO and \CII\ lines (Salom\'e et al. 2012;
Carilli et al. 2013). For the LAEs, the frequency location
of the \NII\ peak emission is marked in Fig.~2, based on 
the \CII\ redshift (Carilli et al. 2013).  Neither LAE is
detected here. 

Both SMG and QSO are spatially resolved in our \NII\ image.
A 2d Gaussian fit to the \NII\ image of the SMG (Fig.~1b)
yielded 2.3\arcsec$\times$1.1\arcsec\ (FWHM along each axis),
with the major axis at PA $\approx 12$\arcdeg.  For the QSO, 
these values are (1.6\arcsec$\times$0.9\arcsec, PA $\approx
91$\arcdeg).  After a deconvolution with the ALMA beam, they
become (2.1\arcsec$\times$0.5\arcsec, PA $\approx 10$\arcdeg)
and (1.3\arcsec$\times$0.4\arcsec, PA $\approx 95$\arcdeg)
for the SMG and QSO, respectively.  The de-convolved major 
axes correspond to $\sim$14 and 9\,kpc for the SMG and QSO, 
respectively.  Both QSO and SMG are unresolved in the continuum
(see \S3.2) and also appear to be unresolved in \CII\ in 
Carilli et al. (2013; beam size 1.2\arcsec$\times$0.8\arcsec). 
Therefore, the \NII\ emission is much more extended (or diffuse)
than the dust or \CII emission in these galaxies. 

For the QSO, the \NII\ channel images reveal that the most 
blue-shifted (red-shifted) emission is located at the south-western
(north-eastern) side of the galaxy, consistent with a rotation.
For the SMG, the \NII\ channel images do not support a 
position-velocity pattern consistent with a rotation. These 
are all consistent with what is seen in CO\,(5$-$4) and
\CII\ (Salom\'e et al. 2012; Carilli et al. 2013).

\subsection{Dust Continuum} \label{sec3.2}

We derive the total continuum flux density of a galaxy by 
fitting a 2d Gaussian to the galaxy image in Fig.~1a after 
temporarily masking out the other galaxies. The resulting
Gaussian FWHMs (in Table 1) confirm that both QSO and SMG 
are unresolved.  [In fact, the insert in Fig. 1a shows that,
even at the enhanced resolution of $\sim$0.7\arcsec, they
are still  unresolved with the following fitted Gaussian
FWHMs: 0.73\arcsec$\times$0.59\arcsec\ (PA $=$ 76\arcdeg) for
the QSO and 0.75\arcsec$\times$0.58\arcsec\ (PA $=$ 89\arcdeg)
for the SMG.]  Table 1 shows that LAE2 is largely resolved.

In Fig.~3 we plot the \CII/CO\,(7$-$6) and \NII/CO\,(7$-$6)
flux ratios, each as a function of $C(60/100)$, for the local
(U)LIRGs and high-$z$ galaxies used in Lu15. The corresponding
flux ratios of the QSO and SMG in BRI 1202-0725 are indicated 
by the horizontal lines.  The black line in each plot is a 
least-squares bisector fit (Isobe et al.~1990) to the local
(U)LIRGs of detections only, excluding the AGNs (see Fig. 3
caption). 
For the SMG, the corresponding line flux ratio intercepts 
the black line at $C(60/100) \approx 1.19$ in (a) and 
$\approx$ 1.23 in (b).  The difference between these values
is smaller than the scatter (0.15 to 0.2) in these plots. 
We therefore simply adopt their average value of 1.21.  
For the QSO, such determined $C(60/100)$ values are 1.22
and 1.15, respectively.  Even though there is some apparent 
segregation between the AGNs and
the rest of the local (U)LIRGs in Fig.~3b, the AGN sample 
size is still too small to draw a firm conclusion on this.  
We therefore also adopt the average $C(60/100) = 1.19$ for 
the QSO.  We set the uncertainty of these FIR colors to 
the horizontal scatter w.r.t.~the black line in the \NII\ 
plot, i.e., $\sim$0.15.  Assuming $\beta = 1.8$ (Planck 
Collaboration et al. 2011), these adopted FIR colors correspond
to $T_{\rm dust} \approx 43$\,($\pm 2$)\,K.

The QSO and SMG are only spatially separated by a few existing
interferometric observations, all at the long-wavelength side of their SED 
peak (see Fig. 4). The $T_{\rm dust}$ determined above provides 
a crucial constraint on the SED shape. Our best modified-blackbody
SED fits for the QSO and SMG separately, after fixing $T_{\rm dust}$
at 43 K and $\beta$ at 1.8, are also shown in Fig. 4.  We note
that similar SED fits would be obtained by using only the continuum 
flux from the CO\,(7$-$6) observation and our adopted $C(60/100)$.  
The luminosity of the SED fit integrated over 20 to 1000\um\ is 
$2.9$ $(3.0) \times 10^{13}\,L_{\odot}$ for the QSO (SMG).
These values are 3-5 times larger than those in Salom\'e et al. (2012).
The \LFIR\ (over 42$-$122\um) from our SED fit is $2.3$ $(2.4) 
\times 10^{13}\,L_{\odot}$, resulting in a $\log L_{\rm CO(7-6)}/L_{\rm FIR} 
\approx -4.54$ ($-4.42$) for the QSO (SMG). These values agree 
with the average of $-4.61$ for our local (U)LIRG sample to 
within $\sim$1.5$\sigma$, where $\sigma$ ($\approx 0.12$) is 
the local sample standard deviation (see Lu15).

\subsection{Star Formation Properties} \label{sec3.3}

Following Lu15, the SFR inferred from $L_{\rm CO(7-6)}$ 
is 5.1 $(6.9) \times 10^3$\,$M_{\odot}\,$yr$^{-1}$ for 
the QSO (SMG) using the initial mass function (IMF) of 
Chabrier (2003).  For local (U)LIRGs, $\Sigma_{\rm SFR}$ 
is empirically correlated with $C(60/100)$ (Liu et al. 
2015) or $T_{\rm dust}$ (Lutz et al. 2016). The scatter
of these correlations is fairly significant, e.g., 
$\sim$0.6 dex in Lutz et al. (2016).  Nevertheless, 
these two independent correlations give comparable estimates
for $\Sigma_{\rm SFR}$: 
$\sim$$2 \times 10^3\,M_{\odot}$\,yr$^{-1}$\,kpc$^{-2}$ for 
both QSO and SMG after we adjusted their correlations 
to Chabrier IMF and increased the \LFIR-based SFR in 
Lutz \etal by a factor of 2 to align with the Kennicutt (1998)
formula. These estimates of $\Sigma_{\rm SFR}$ are quite 
high, but still below the Eddington limit of $\sim$$3 \times 
10^3\,M_{\odot}$\,yr$^{-1}$\,kpc$^{-2}$ (Murray et al. 2005; 
Thompson et al. 2005; Hopkins et al. 2010).  The face-on 
FWHM diameter, $d$, of the SF region can be estimated via 
$\Sigma_{\rm SFR} = (\frac{1}{2}\,{\rm SFR})/({\frac{1}{4}}\pi\,d^2)$. 
For $\Sigma_{\rm SFR} = 2 \times 10^3\,M_{\odot}$\,yr$^{-1}$\,kpc$^{-2}$, 
the resulting $d = 1.3$ (1.5)\,kpc for the QSO (SMG), 
consistent with them being unresolved in our continuum image.
These sizes are somewhat smaller than the $\sim$2\,kpc scale 
resolved in a high-resolution CO\,(2$-1$) image (Carilli 
et al. 2002), a phenomenon that is also seen locally (Xu et 
al. 2014, 2015).  Following Scoville et al. (2016), we 
estimated the molecular gas mass, $M_{\rm gas}$, to be 
$5 \times 10^{11}\,M_{\odot}$ for either galaxy based on 
the rest-frame $f_{\nu}$(850\um) from our SED fit. The formal
uncertainty for $M_{\rm gas}$ is about a factor of 2.
The gas depletion time $\tau_{\rm gas} \approx $ 10 (7) 
$\times 10^7$\,years for the QSO (SMG).  

\subsection{On {\rm \NII/\CII}\ Ratios} \label{sec3.4}

For local (U)LIRGs, $\log\,L_{\rm [NII]}/L_{\rm [CII]}$ 
correlates linearly with $C(60/100)$, with a scatter of
only $\sim$0.15 in $\log\,L_{\rm [NII]}/L_{\rm [CII]}$ (Lu15).  
As $C(60/100)$ increases from 0.4 to $\sim$1.3,  
$L_{\rm [NII]}/L_{\rm [CII]}$ drops by a factor of 4 (from 
0.12 to 0.03).  This correlation is due to an increasing 
contribution to \CII\ from the neutral medium around young 
massive stars as $C(60/100)$ increases (see Da\'iz-Santos 
et al. 2017).  Our observation shows that both QSO and SMG 
in BRI 1202-0725 follow this local 
trend as well.  

Theoretical considerations predict a smaller \NII/\CII\ ratio
for a lower metallicity (Nagao et al 2012; Pereira-Santaella
et al. 2017). In such a scenario, galaxies of different 
metallicities follow separate \NII/\CII$-$$C(60/100)$ tracks.  
In principle, one needs to know both \NII/\CII\ and $C(60/100)$
in order to constrain the metallicity of a galaxy.  For example,
our current observation sets \NII/\CII\ $< 0.03$ for the two
LAEs.  These limits alone are not stringent enough to conclude
if the LAEs follow a different \NII/\CII$-$$C(60/100)$ track 
than the local (U)LIRGs.

\acknowledgments

This paper benefited from a number of thoughtful comments 
made by the anonymous referee.  This paper makes use of 
the following ALMA data: ADS/JAO.ALMA\#2015.1.00388.S. 
ALMA is a partnership of ESO (representing its member 
states), NSF (USA) and NINS (Japan), together with NRC 
(Canada), NSC and ASIAA (Taiwan), and KASI (Republic of 
Korea), in cooperation with the Republic of Chile. 
The Joint ALMA Observatory is operated by ESO, AUI/NRAO
and NAOJ. This work is supported in part by the NSFC 
grants 11673028, 11673057 and 11420101002. TDS, GCP and 
DR acknowledge CONICYT project \#31130005 and FONDECYT
project \#1151239, FONDECYT fellowship \#3150361, and 
grant ST/N000919/1, respectively.

\vspace{0.25in}


\newpage


\begin{deluxetable*}{llllll}
\tabletypesize{\footnotesize}
\tablenum{1}
\tablewidth{0pt}
\tablecaption{BRI\,1202-0725: Observed and Derived Parameters$^a$}
\tablehead{
\colhead{\hspace{-0.75in}Parameter}       & \colhead{\hspace{-1.0in}Unit}      & \colhead{\hspace{-0.4in}QSO} 	     & \colhead{\hspace{-0.4in}SMG}     & \colhead{\hspace{-0.4in}LAE2}     & \colhead{\hspace{-0.1in}LAE1}}
Continuum: \\
\ \ \ R.A.$^b$                  & \hspace{-0.25in}J2000              & 12$^{\rm h}$05$^{\rm m}$23{\fs}13              & 12$^{\rm h}$05$^{\rm m}$22{\fs}98       & 12$^{\rm h}$05$^{\rm m}$23{\fs}04       & \nodata \ \ \ \ \\
\ \ \ Decl.$^b$	                & \hspace{-0.25in}J2000              & -7{\arcdeg}42{\arcmin}32{\farcs}8              & -7{\arcdeg}42{\arcmin}29{\farcs}7       &  -7{\arcdeg}42{\arcmin}34{\farcs}5      & \nodata \\
\ \ \ $S_{\nu}$(256 GHz)$^c$    & \hspace{-0.25in}mJy          & 7.58 ($\pm 0.01$)           & 7.20 ($\pm 0.01$)           & 0.83 $\pm 0.01$           & $<0.10$  \\
\ \ \ Gaussian fit$^c$         & \hspace{-0.25in}\nodata      & (1.08\arcsec$\times$0.89\arcsec, 73\arcdeg)  & (1.08\arcsec$\times$0.87\arcsec, 75\arcdeg)  & (1.20\arcsec$\times$0.98\arcsec, 122\arcdeg)  & \nodata \\

\\
\NII: \\
\ \ \ Aperture$^d$	        & \hspace{-0.25in}\nodata            & (3\arcsec$\times$1.65\arcsec, 90\arcdeg)  & (2.65\arcsec$\times$1.6\arcsec, 0\arcdeg) & (1.6\arcsec$\times$1.6\arcsec, 0\arcdeg)    & (1\arcsec$\times$1\arcsec, 0\arcdeg) \\
\ \ \ $z^e$		        & \hspace{-0.25in}\nodata            & 4.695                       & 4.693               & \nodata                     & \nodata \\
\ \ \ Flux$^f$                  & \hspace{-0.25in}Jy\kms             & 1.01 ($\pm 0.02$)           & 0.99 ($\pm 0.02$)   & $<0.13$                     & $<0.02$ \\
\ \ \ FWHM$^g$                  & \hspace{-0.25in}km\,s$^{-1}$       & 380                         & 794                 & \nodata                     & \nodata \\ 
\ \ \ Diameter$^h$	          & \hspace{-0.25in}kpc	             & 9                    	   & 14			 & \nodata		       & \nodata \\
\\
Derived parameters:\\
\ \ \ $L_{CO(7-6)}^i$           & \hspace{-0.25in}$10^8\,L_{\odot}$                            & 6.6 ($\pm 0.6$)      &  8.9 ($\pm 1.2$)    & \nodata     & \nodata \\
\ \ \ $L_{\rm [CII]}^j$         & \hspace{-0.25in}$10^9\,L_{\odot}$                            & 6.5 ($\pm 1.0$)      & 10.0 ($\pm 1.5$)    & \nodata     & \nodata \\
\ \ \ $L_{\rm [NII]}$           & \hspace{-0.25in}$10^8\,L_{\odot}$                            & 5.3 ($\pm 0.1$)      & 5.2 ($\pm 0.1$)     & \nodata     & \nodata \\
\ \ \ $T_{\rm dust}^k$          & \hspace{-0.25in}K                                            & 43 ($\pm 2$)         & 43 ($\pm 2$)        & \nodata     & \nodata \\
\ \ \ $L_{\rm FIR}^l$           & \hspace{-0.25in}$10^{13}\,L_{\odot}$                         & 2.3 ($\pm 18\%$)     & 2.4 ($\pm 18\%$)    & \nodata     & \nodata \\
\ \ \ $L_{\rm IR}^m$            & \hspace{-0.25in}$10^{13}\,L_{\odot}$                         & 5.0 ($\pm 30\%$)     & 6.7 ($\pm 30\%$)    & \nodata     & \nodata \\
\ \ \ SFR$^m$                   & \hspace{-0.25in}$10^3\,M_{\odot}$\,yr$^{-1}$                 & 5.1 ($\pm 30\%$)     & 6.9 ($\pm30\%$)	    & \nodata     & \nodata \\
\ \ \ $\Sigma_{\rm SFR}^n$      & \hspace{-0.25in}$10^2\,M_{\odot}$\,yr$^{-1}$\,kpc$^{-2}$      & 16/21       & 18/21             & \nodata	    & \nodata \\
\ \ \ SF diameter$^o$           & \hspace{-0.25in}kpc                                           & 1.3         & 1.5	          & \nodata	    & \nodata \\
\ \ \ $M_{\rm H_2}^p$           & \hspace{-0.25in}$10^{11}\,M_{\odot}$                          & 5           & 5                 & \nodata	    & \nodata \\
\ \ \ $\tau_{\rm gas}^q$        & \hspace{-0.25in}$10^7$ years                                  & 10	      & 7                 & \nodata	    & \nodata \\
\enddata
\tablenotetext{a}{All SFRs are consistent with eq.~(4) in Kennicutt (1998), but scaled to Chabrier IMF.  The ALMA flux uncertainties cited do not include 
		  the absolute calibration uncertainty.}
\tablenotetext{b}{From the Gaussian fit.}
\tablenotetext{c}{Continuum flux density from the Gaussian fit of which the major and minor FWHM axes and the major axis PA are given.
                  The 3$\sigma$ flux upper limit for LAE1 assumes an unresolved case. }
\tablenotetext{d}{Aperture for the 1d spectrum extraction:  major $\times$ minor axes, followed by the major axis PA.}
\tablenotetext{e}{Redshift (in LSR) from the peak frequency of the Gaussian spectral line fit.}
\tablenotetext{f}{Flux from the Gaussian line fit. The upper limit for LAE2 was derived using a Gaussian profile 
		     with a peak flux density of 3$\sigma$ and a FWHM equal to that of the \CII\ emission in Carilli et al (2013). 
		     For LAE1, which has a narrow line width, this is simply 3$\sigma$ times 100\,\kms.  Here $\sigma$ is 
	             the channel-to-channel noise in the spectrum in Fig. 2.}
\tablenotetext{g}{FWHM of the Gaussian line fit.}
\tablenotetext{h}{Major axis of the Gaussian fit to the \NII\ image in Fig. 1, after deconvolution with the ALMA beam.}
\tablenotetext{i}{Based on the line fluxes in Salom\'e et al. (2012).}
\tablenotetext{j}{Based on the line fluxes in Wagg et al. (2012).}
\tablenotetext{k}{Adopted dust temperature.}

\tablenotetext{l}{The 42-122\um\ luminosity from the FIR dust SED fit. The uncertainty corresponds to a variation of 
		     $\pm$2\,K in $T_{\rm dust}$.}
\tablenotetext{m}{Inferred from $L_{\rm CO(7-6)}$.}
\tablenotetext{n}{Two estimates on $\Sigma_{\rm SFR}$: the first one from $C(60/100)$ (Liu et al. 2015) and 
		     the second one from the 70-to-160\um\ color of the SED fit (Lutz et al. 2016).}
\tablenotetext{o}{FWHM of the SF region, assuming $\Sigma_{\rm SFR} = 2,000\,M_{\odot}$\,yr$^{-1}$\,kpc$^{-2}$.}
\tablenotetext{p}{Molecular gas mass inferred from the rest-frame 850\um\ luminosity of the SED fit.}
\tablenotetext{q}{Gas depletion time.}
\end{deluxetable*}

\newpage

\null\vspace{5in}
\begin{figure*}
\centering
\includegraphics[width=1.0\textwidth, bb=0 600 930 1120]{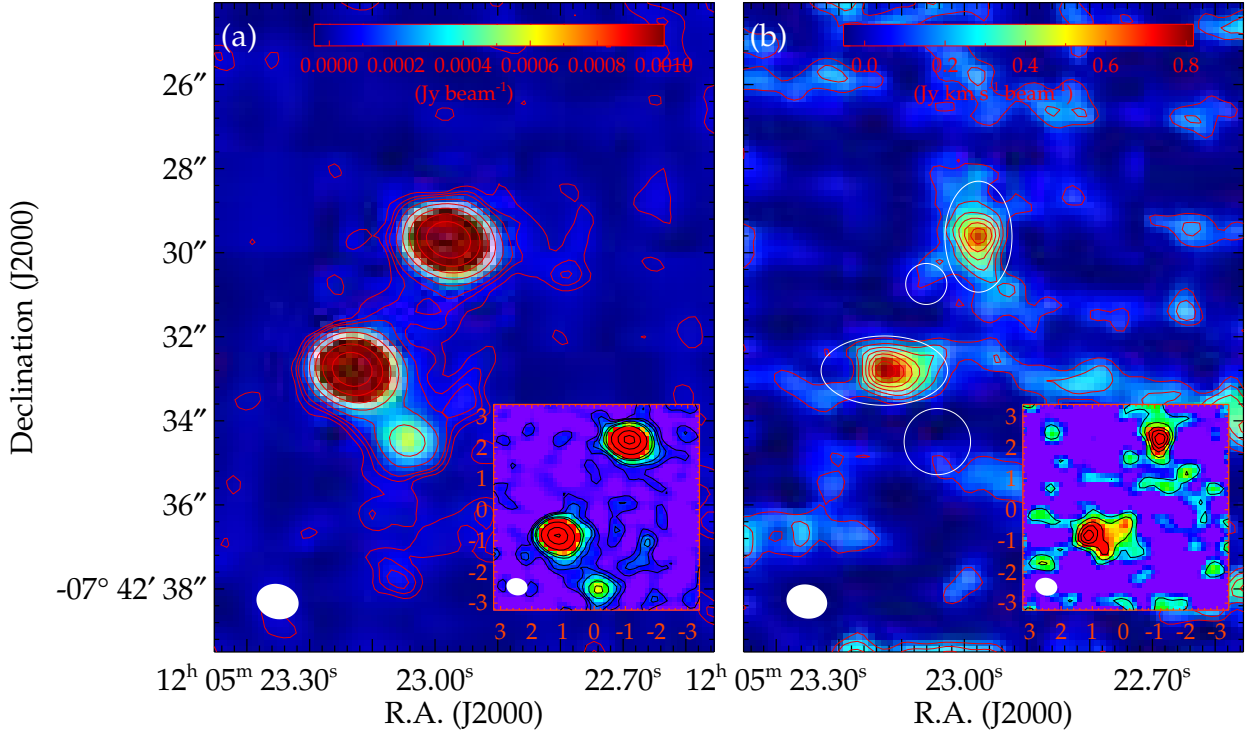}
\vspace{1.0in}
\caption{
Images of (a) the continuum in Jy\,beam$^{-1}$ and (b) \NII\ in Jy\,\kms\,beam$^{-1}$,
using the data cleaned with the natural weighting. The effective beam 
(1.0\arcsec$\times$0.8\arcsec, PA $=$ 75\arcdeg) is shown near the bottom left 
corner.  The \NII\ image was integrated over $\nu_{\rm obs} = 256.148$ to 257.090 GHz.
Both image and contours refer to the same emission.  The continuum and \NII\ contours
are at (1, 2, 3, 6, 12, 24, 48, 96, 147, 176.5) $\times 34\,\mu$Jy\,beam$^{-1}$ 
(= 1$\sigma$) and (1, 2, 3, 4, 5, 7) $\times 0.104$\,Jy\,\kms\,beam$^{-1}$ (= 1$\sigma$),
respectively.  The 4 white ellipses in the \NII\ image mark the apertures used to 
extract the 1d spectra.  The smallest ellipse is centered on LAE1. The insert 
in each panel shows a 6.75\arcsec$\times$6.75\arcsec\ section, centered at 
(12$^{\rm h}$05$^{\rm m}$23{\fs}05, -7{\arcdeg}42{\arcmin}31{\farcs}85), of 
the same image, but based on the data cleaned with the Briggs weighting (with 
{\it robust} = 0) that produced a finer effective beam of $\sim$0.7\arcsec$\times$0.6\arcsec\
(as plotted).  Here the contours are at (1, 2, 3, 6.3, 9.4, 25, 62.5, 104) $\times 
48\,\mu$Jy\,beam$^{-1}$ (= 1$\sigma$) and (1, 2, 3, 3.9, 4.7) $\times 
0.117$\,Jy\,\kms\,beam$^{-1}$ (= 1$\sigma$) for the continuum and \NII\ images,
respectively.
}
\label{Fig1}
\end{figure*}

\vspace{5.0in}
\newpage

\begin{figure*}
\begin{center}
\begin{tabular}{cc}
\includegraphics[width=0.40\textwidth, bb=250 300 650 780]{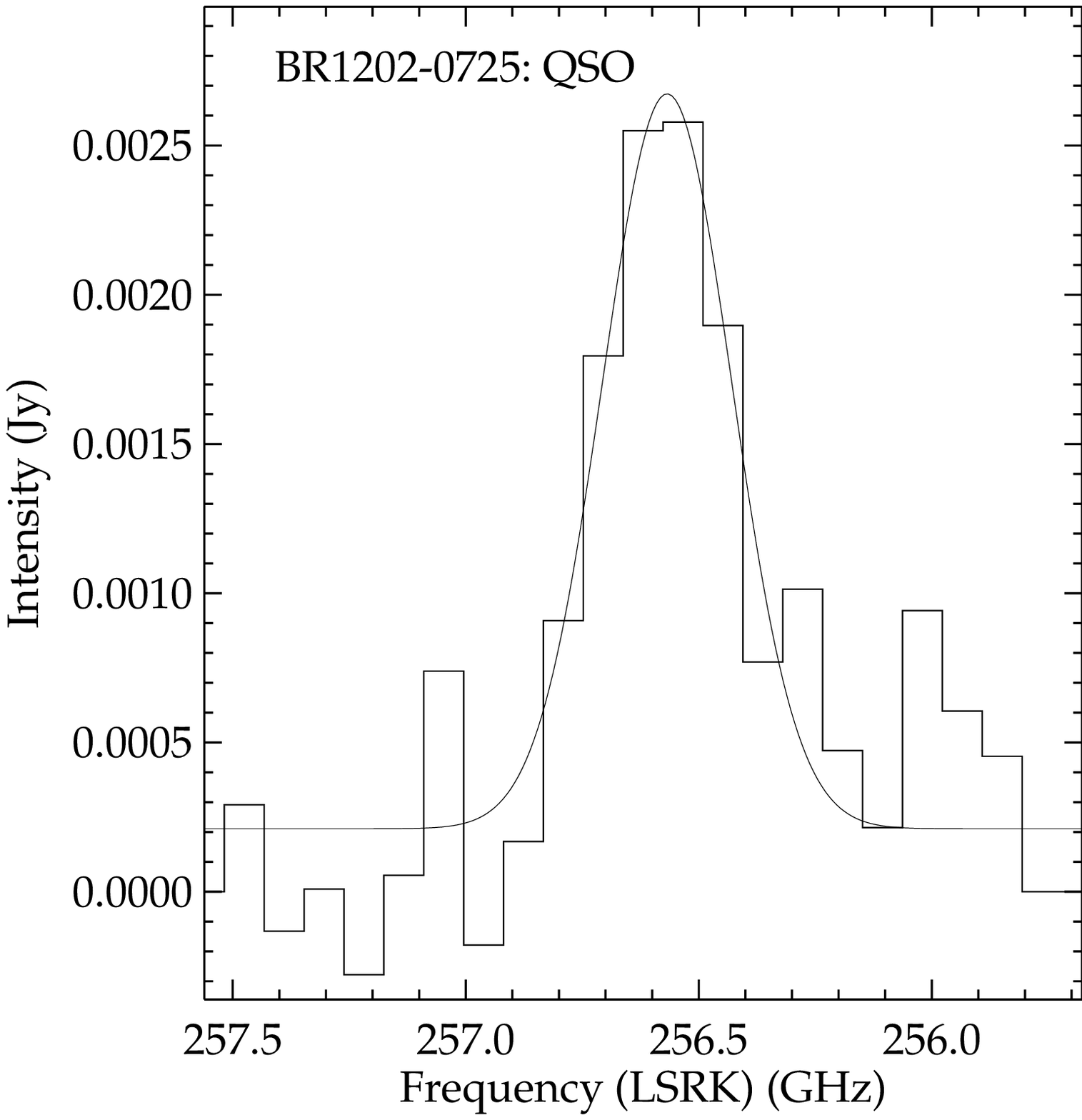} & \includegraphics[width=0.40\textwidth, bb=200 300 600 780]{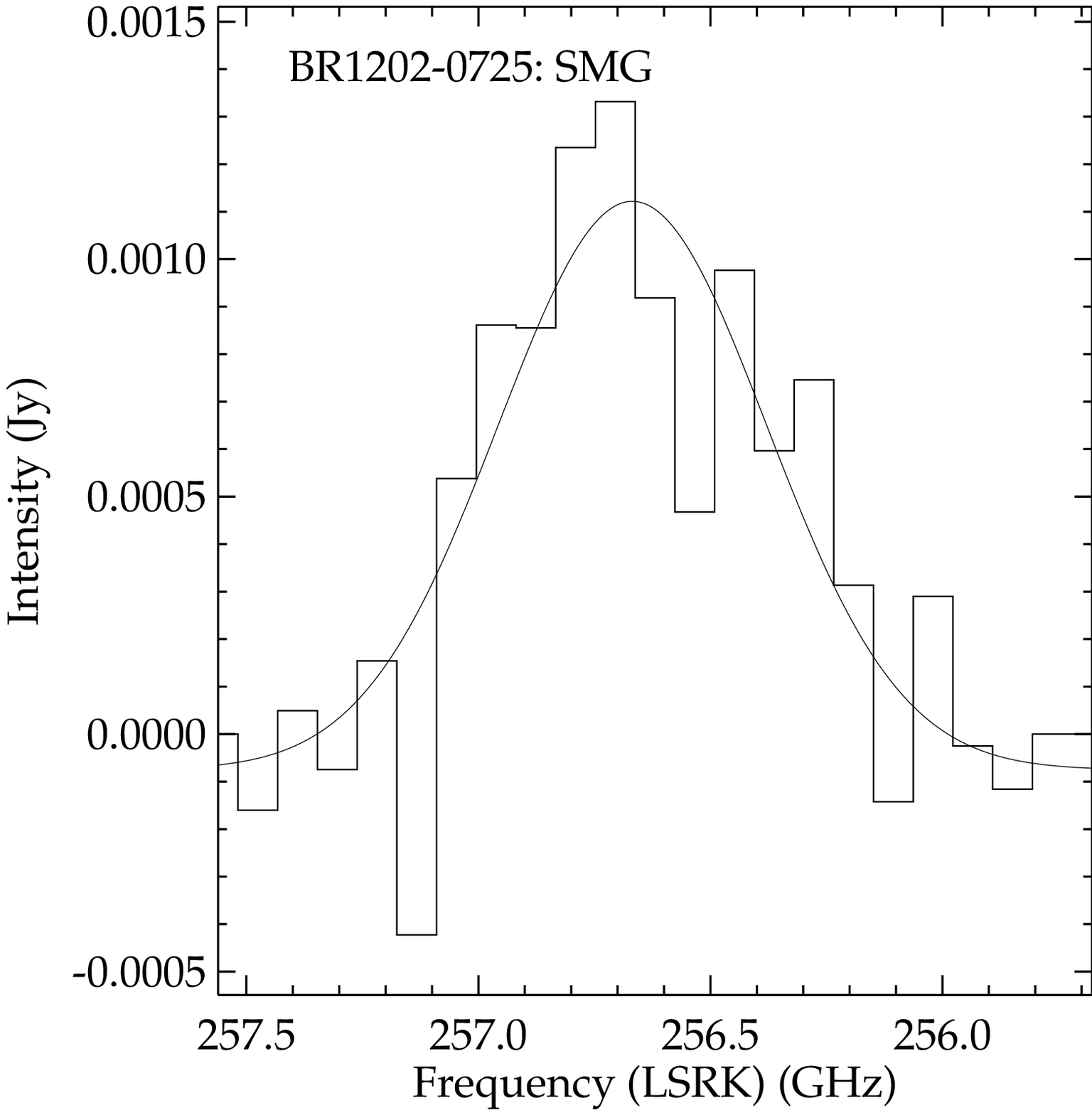} \\
\includegraphics[width=0.40\textwidth, bb=250 300 650 780]{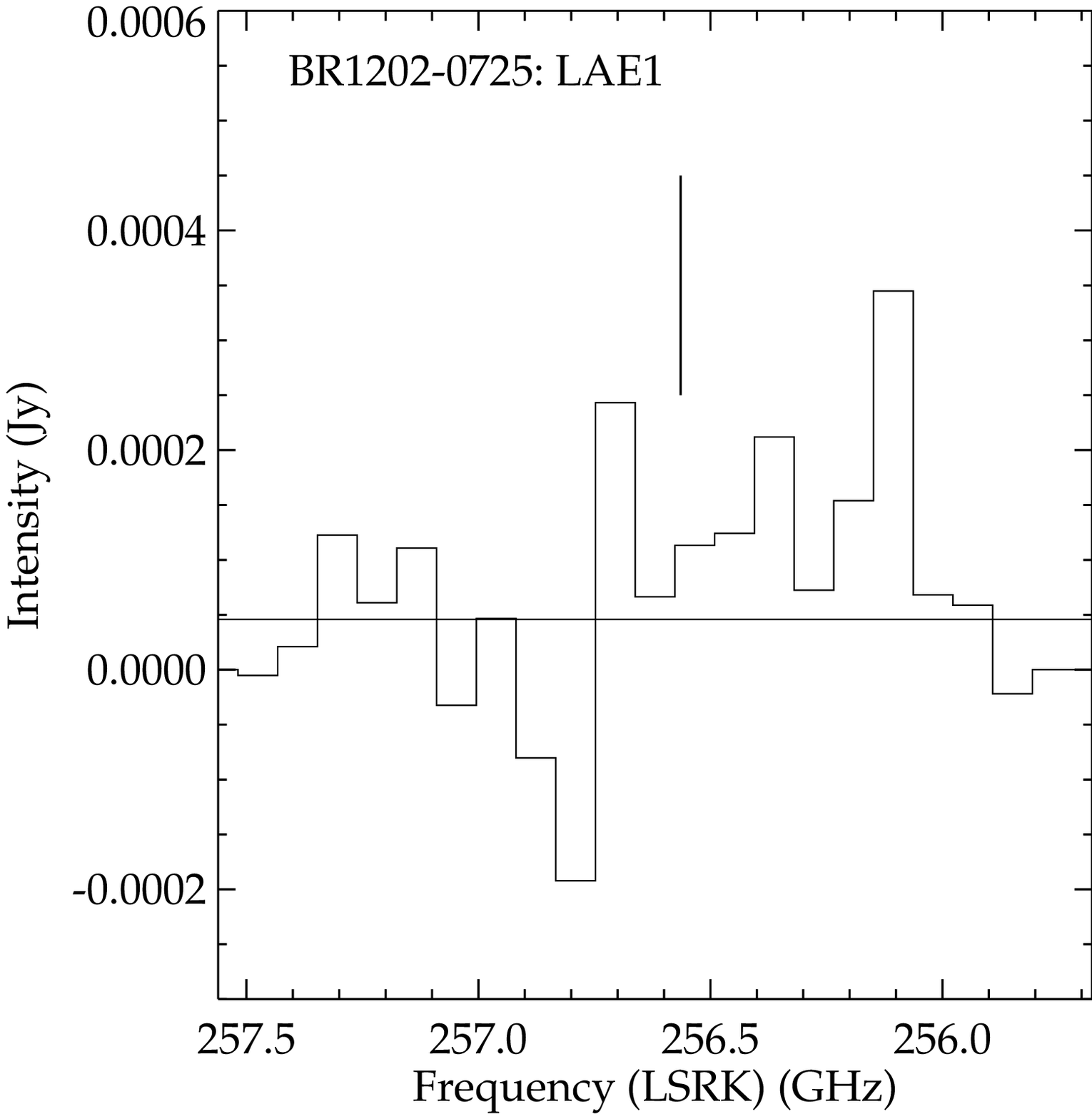} & \includegraphics[width=0.40\textwidth, bb=200 300 600 780]{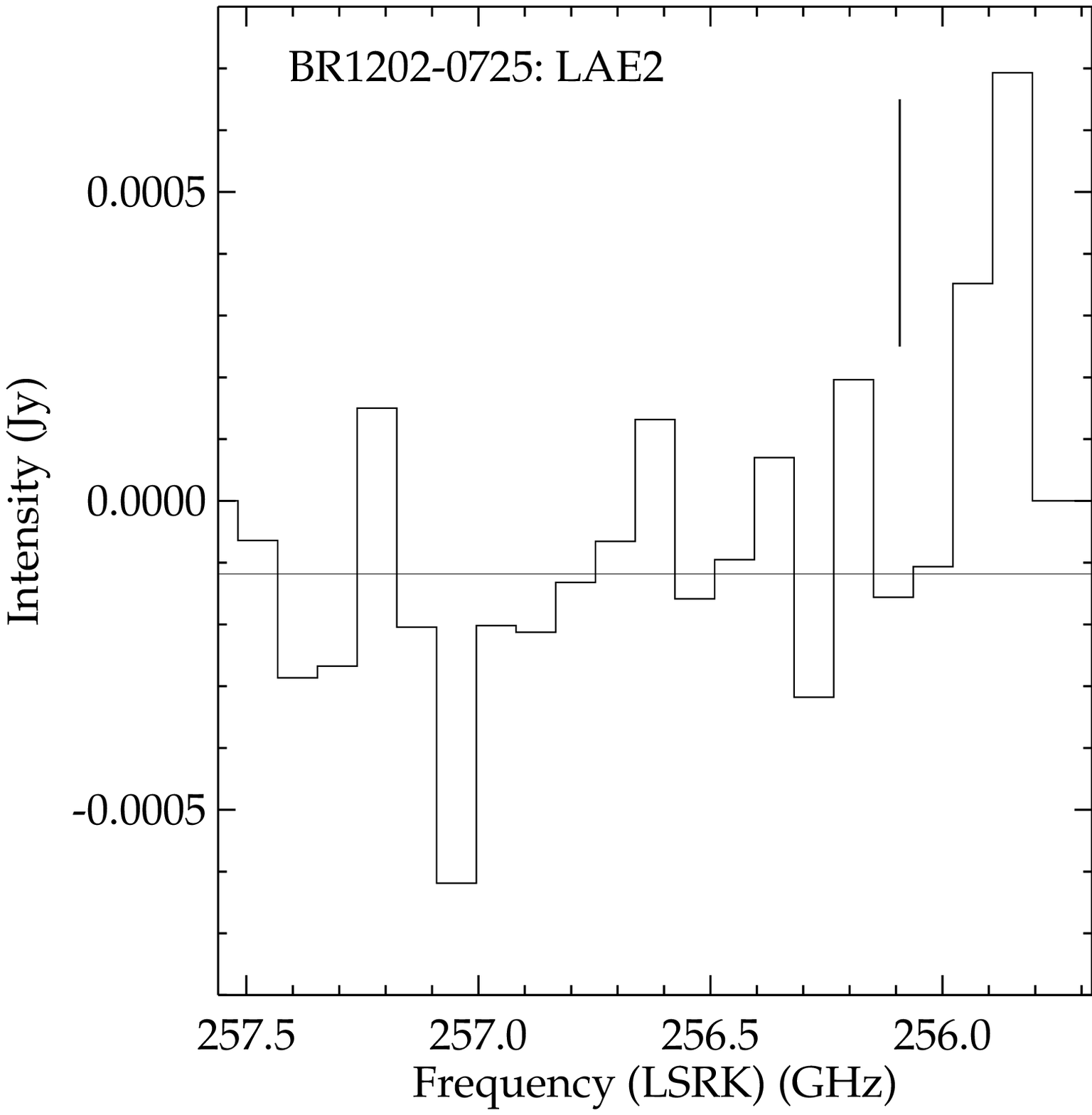} \\
\end{tabular}
\end{center}
\vspace{0.8in}
\caption{
Individual spectra extracted using the elliptical apertures shown in Fig.~1,
plotted as a function of the observed frequency. Each frequency bin corresponds
to 100\kms.   The Gaussian fits are also plotted for both QSO and SMG.  The vertical
bar in the spectra of LAE1 and LAE2 indicates the \NII\ peak frequency expected
from the redshift of the \CII\ line observed in Carilli et al. (2013).
}
\label{Fig2}
\end{figure*}

\vspace{5.0in}
\newpage

\begin{figure*}
\centering
\includegraphics[width=.95\textwidth, bb=18 144 592 718]{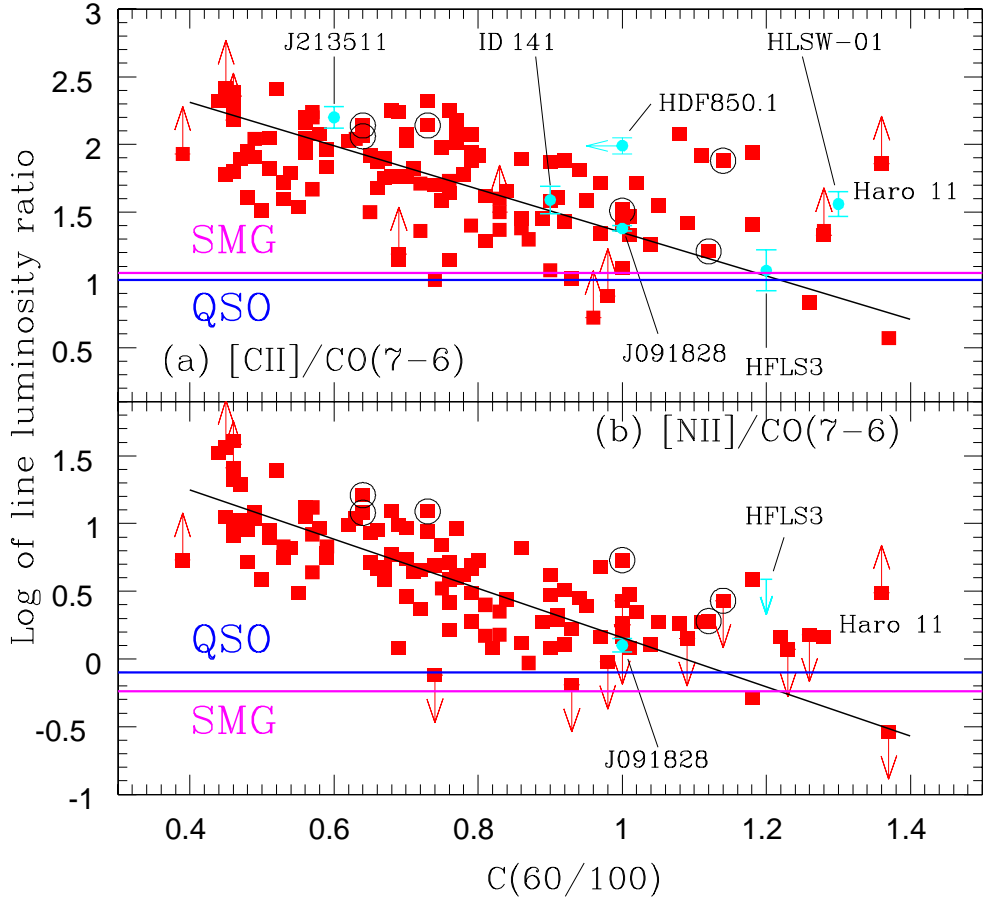}
\caption{
Comparisons of the QSO (blue line) and SMG (magenta line) in BRI\,1202-0725 
with (i) the local (U)LIRGs (red squares) and (ii) high-$z$ galaxies from Lu15
in plots of (a) the logarithmic \CII\ to CO\,(7$-$6) luminosity ratio and (b) 
the logarithmic \NII\ to CO\,(7$-$6) luminosity ratio, as functions of 
the FIR color.  The black solid line in each plot is a least-squares bisector 
fit to the detections of the local (U)LIRGs, except for the 6 dominant AGNs 
(further circled; see Lu15).  For the \NII\ line, this fit is given in eq. (6) 
in Lu15;  for the \CII\ line, this fit is: $\log\,$\CII/CO\,(7$-$6) $= (-1.61 
\pm 0.12)\,C(60/100) + (2.95 \pm 0.09)$.
}
\vspace{0.5in}
\label{Fig3}
\end{figure*}

\vspace{5.0in}
\newpage

\begin{figure*}
\centering
\includegraphics[width=.95\textwidth, bb= 18 144 592 718]{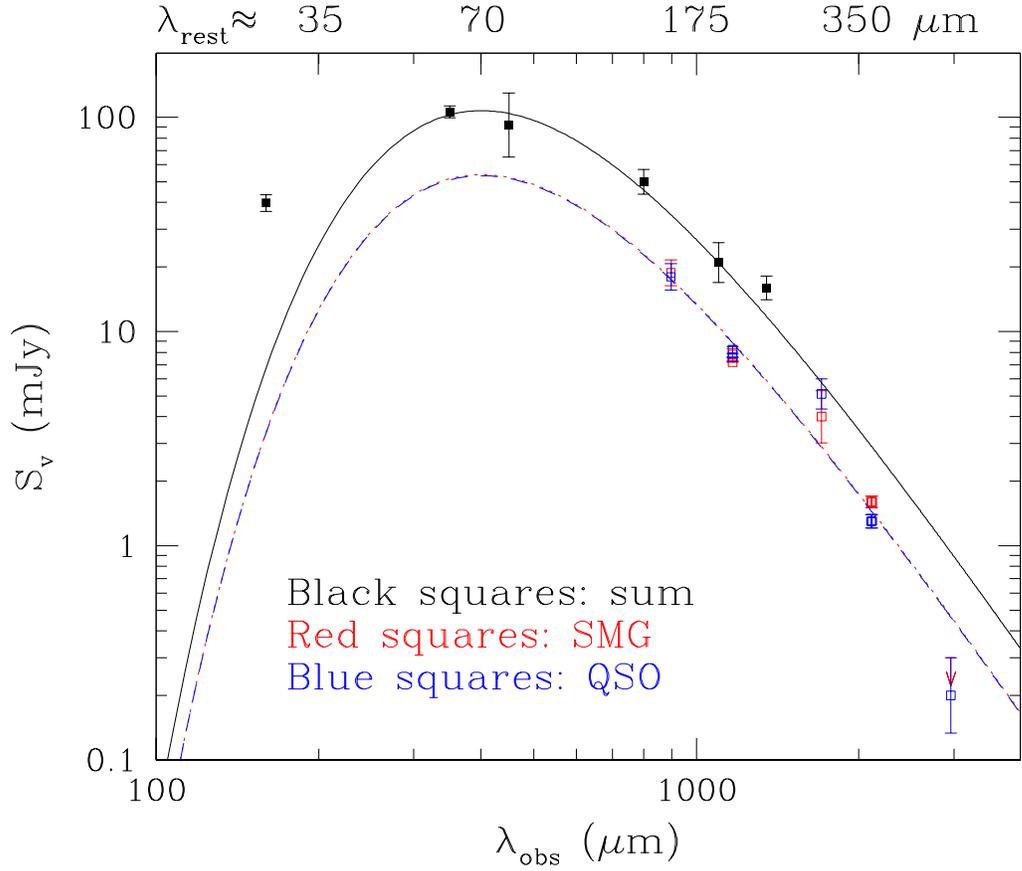}
\caption{
Plot of the continuum measurements (squares) and model SED fits (dashed curves) as a 
function of $\lambda_{\rm obs}$ for the QSO (in blue) and SMG (red). The measurements 
applicable only to the two galaxies combined are shown in black, so is the sum of 
the individual SED fits.  Each SED fit fixed $\beta = 1.8$ and $T_{\rm dust} = 43$\,K.  
The data points in color are at $\lambda_{\rm obs} =$ 2961\um, 2110\um, 1705\um\ 
(Salom\'e et al. 2012), 1167\um\ (this work and  Decarli et al. 2014) and 898\um\ 
(Wagg et al. 2012), respectively.  The black data points are at $\lambda_{\rm obs} 
= 1350$\um\ (Omont et al. 1996), 1100\um, 800\um, 450\um\ (Isaak et al. 1994), 350\um\ 
(Benford et al. 1999) and 160\um\ (Leipski et al. 2010), respectively.}
\vspace{0.5in}
\label{Fig4}
\end{figure*}

\vspace{5.0in}
\newpage

\end{document}